\documentclass[conference]{IEEEtran}
\usepackage{array}
\usepackage{graphicx}
\usepackage{amsmath}
\usepackage{algorithmic}
\usepackage{algorithm}
\usepackage{epsfig}
\usepackage{url}
\usepackage{epstopdf}
\usepackage{cite}
\usepackage[font={small},skip=5pt]{caption}
\usepackage{subcaption}
\usepackage{setspace}
\usepackage{enumitem}
\usepackage{graphicx,lipsum}
\usepackage{caption}
\usepackage{tikz}
\usepackage{lipsum}

\newcolumntype{P}[1]{>{\centering\arraybackslash}p{#1}} 
\newcolumntype{M}[1]{>{\centering\arraybackslash}m{#1}} 
\newcommand\copyrighttext{%
  \footnotesize \copyright~2015 IEEE. Personal use of this material is permitted. Permission from IEEE must be obtained for all other uses, in any current or future media, including reprinting/republishing this material for advertising or promotional purposes, creating new collective works, for resale or redistribution to servers or lists, or reuse of any copyrighted component of this work in other works.}
\newcommand\copyrightnotice{%
\begin{tikzpicture}[remember picture,overlay]
\node[anchor=south,yshift=10pt] at (current page.south) {\fbox{\parbox{\dimexpr\textwidth-\fboxsep-\fboxrule\relax}{\copyrighttext}}};
\end{tikzpicture}%
}

\begin{document}


\bibliographystyle{IEEEtran}
%

\title{Predicting Performance of Channel Assignments in Wireless Mesh Networks through Statistical Interference Estimation}

\author{\IEEEauthorblockN{Srikant Manas Kala, M Pavan Kumar Reddy, and Bheemarjuna Reddy Tamma}
\IEEEauthorblockA{ Indian Institute of Technology Hyderabad, India\\
Email: [cs12m1012, cs12b1025, tbr]@iith.ac.in}}

\maketitle

\begin{abstract}
Wireless Mesh Network (WMN) deployments are poised to reduce the reliance on wired infrastructure especially with the advent of the multi-radio multi-channel (MRMC) WMN  architecture. But the benefits that MRMC WMNs offer \emph{viz.}, augmented network capacity, uninterrupted connectivity and reduced latency, are depreciated by the detrimental effect of prevalent interference. Interference mitigation is thus a prime objective in WMN deployments. It is often accomplished through prudent channel allocation (CA) schemes which minimize the adverse impact of  interference and enhance the network performance. However, a multitude of CA schemes have been proposed in research literature and absence of a CA performance prediction metric, which could aid in the selection of an efficient CA scheme for a given WMN, is often felt. 
In this work, we offer a fresh characterization of the interference endemic in wireless networks. We then propose a reliable CA performance prediction metric, which employs a statistical interference estimation approach. We carry out a rigorous quantitative assessment of the proposed metric by validating its CA performance predictions with experimental results, recorded from extensive simulations run on an ns-3 802.11g environment.
\end{abstract}

\section{Introduction}

Wireless Mesh Networks (WMNs) have sparked a great interest in the research community as they offer reliable connectivity coupled with remarkably enhanced bandwidths. The most effective WMN framework is the multi-radio multi-channel (MRMC) deployment, which harnesses the availability of several non-overlapping channels under the IEEE 802.11 and IEEE 802.16 standards. However, the advent of WMN technology has also spawned a plethora of performance related issues in WMNs, which include the problems of channel allocation to radios, routing, scheduling etc. At the core of these issues lies the impeding factor of interference, which is caused and experienced by concurrently transmitting radios, operating on the same channel and located within each other's interference range. 
Consequently, interference is the single most debilitating factor in WMN performance and substantial research effort is focused on mitigating and restraining its adverse impact. Minimizing interference in a WMN is a primary network design consideration often achieved through a prudent channel assignment (CA) to the radios in the WMN. The CA problem is an NP-Hard problem \cite{NPcomplete} and numerous CA schemes have been proposed in prior research studies \cite{Ding} which strive to alleviate the impact of interference through a variety of innovative approaches.
  
\section{Motivation and Related Research Work}
 
The measure of the degradation of network performance by interference in a WMN deployment is intricately linked to the channel assignment (CA) scheme being employed. There is a multitude of CA schemes which can be implemented in any WMN deployment. However, selecting the most efficient and feasible CA from the enormous set of all CAs, for a given WMN of certain architecture and topology, is a tedious task. Further, there is an absence of CA performance prediction or estimation techniques in the research literature that could aid a network administrator in making this crucial choice. The conventional approach of estimating impact of interference in a WMN is to compute the total interference degree or \textit{TID} \cite{TID1}, which equals half the sum of \textit{interference degrees} of all the links in the graph representing a WMN. Here interference degree denotes the number of links that may potentially interfere or conflict with a given link. A TID estimate accounts for every potential conflict 
link in the WMN and is generated through its \textit{conflict graph}. Thus, the TID estimate is an approximate measure of the endemic interference, and its magnitude is often assumed to be a reliable measure of the adverse impact of interference. For example, in interference-aware CA schemes, the guiding idea is to lower the TID to reduce the intensity of prevalent interference and make the CA more efficient \cite{Arunabha}. However, in a recent study \cite{Manas}, we have demonstrated that although TID does give a measure of the impact of interference, it is inconsistent and unreliable as a CA performance prediction, or interference estimation metric. Thus, we can not, with high confidence, compare two CAs or select the most efficient CA for a given WMN from a set of CA schemes by employing their TID estimates as the sole criterion for performance prediction. Secondly, TID estimation is computationally expensive.

In this work, we aim to remedy this problem by first proposing a fresh theoretical characterization of the interference prevalent in WMNs. We then suggest an intuitive statistical \textit{interference estimation} or \textit{CA performance prediction} technique, based on the proposed classification. Through extensive simulations we demonstrate that the proposed metric is more reliable than the conventional TID metric. In addition, the proposed metric also consumes lesser computational resources.
%
\copyrightnotice
  
\section{Interference In WMNs : A Fresh Characterization}
\begin{figure}[htb!]
                \centering
                \includegraphics[width=8.5cm, height=6cm]{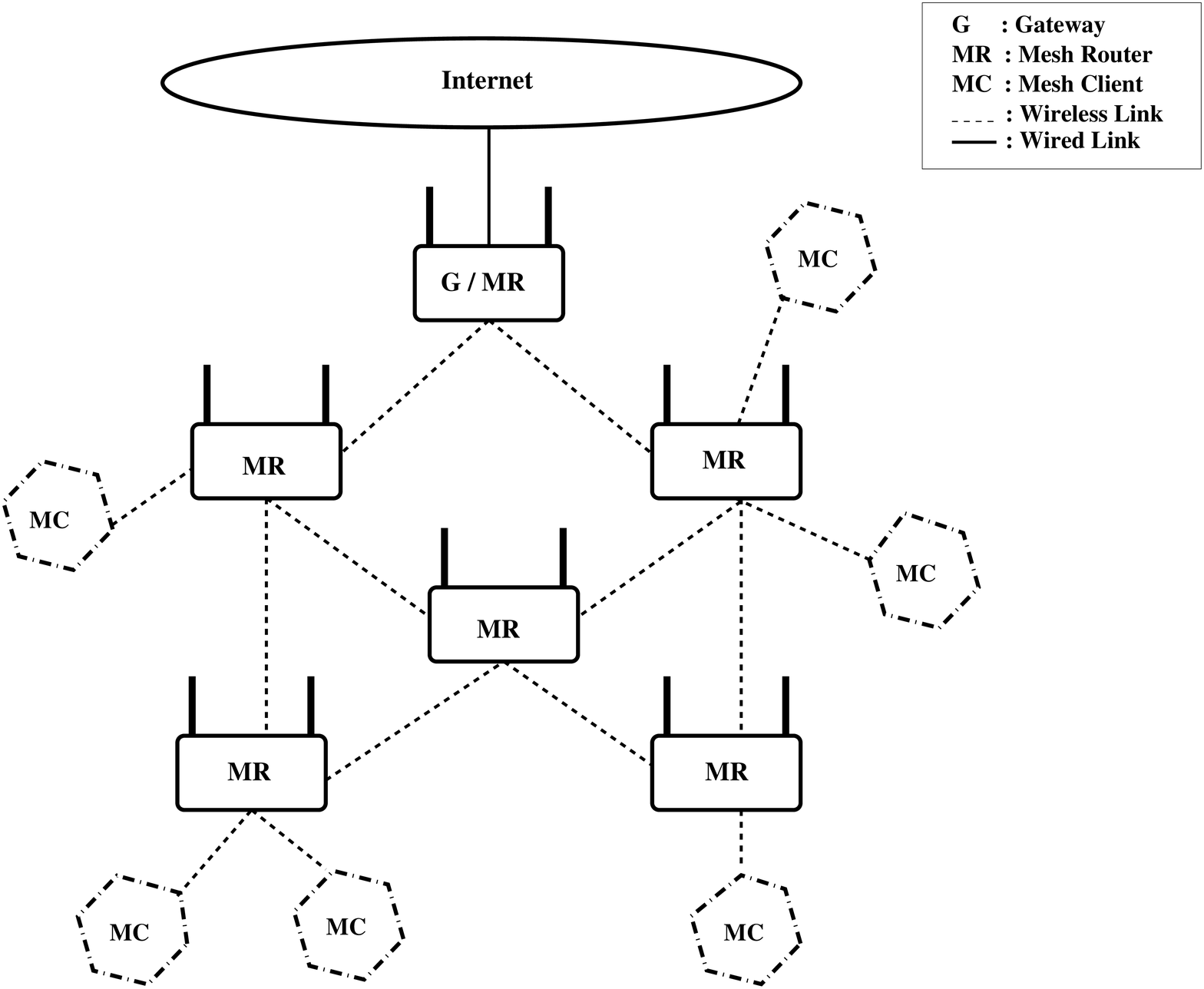}
                \caption{WMN Architecture For Interference Characterization}
                \label{WMN}
        \end{figure}

The classification of interference in wireless networks is based on the \textit{source} of the conflicting wireless transmissions on an identical channel. Thus, it is categorized into \textit{internal}, \textit{external} and \textit{multi-path fading} \cite{26Koutsonikolas}. However, this view is rather simplistic and fails to reflect the inherent \textit{characteristics} which are intrinsic to all forms of interference affecting the wireless communication, regardless of their source or cause. We adopt a fresh approach to elicit these inherent characteristics of interference as they aid us in assessing and estimating its adverse impact on the network performance.
We consider a single-gateway WMN model depicted in \mbox{Figure \ref{WMN}}, which comprises of mesh-routers (nodes) and mesh-clients. Multiple radios are available for inter mesh-router communication and we focus on the interference characterization of the mesh backbone.
        
Next, to study these characteristics we consider a network scenario in which all nodes are active participants of multiple concurrent wireless transmissions. Each node functions as a source, or a destination, or an intermediate node relaying the data packets onto the  next hop. As all nodes transmit in tandem, they trigger and intensify the intricate interference bottlenecks in a WMN, fashioning a prefect scenario in which interference can be considered to be a three dimensional entity, the dimensions being \textit{temporal, spatial and statistical}. We now deliberate over this three dimensional model which constitutes the set of characteristics of the endemic interference. 
\begin{enumerate}
 \item \textbf{Temporal Characteristics} : These characteristics represent the \textit{dynamism} in the interference scenarios. In a wireless network the transmissions are seldom synchronized, and on the contrary, are quite random. Thus, the interference complexities that are spawned in the network are a function of time and fundamentally temporal. 
 \item \textbf{Spatial Characteristics} : Link conflicts in a WMN are a result of two or more interfering links which are in close proximity. The links emanating from two radios transmitting on an identical channel would interfere \textit{if and only if} they lie within each other's \textit{interference range}. Consequently, this spatial interaction of wireless links is a fundamental feature of endemic interference.
 \item \textbf{Statistical Characteristics} : The complexity of interference in a WMN is intricately linked to the assignment of available channels to the radios in the WMN. An even and judicious distribution of channels among radios will spawn fewer wireless conflicts as compared to a skewed distribution. 
\end{enumerate}

\section{Interference Estimation}
Interference estimation, alignment and cancellation are established NP-Hard problems \cite{Alignment}. A theoretical estimate of interference is only an approximate prediction of the WMN performance under a particular CA scheme. It helps to avoid the time and resource consuming task of ascertaining CA performance by implementing a CA in the WMN and carrying out real-time assessments. A TID estimate is the commonly used measure of the impact of endemic interference on WMN performance. The TID metric only factors in the spatial aspects of interference by generating an estimate of the link conflicts, and does not take into account the other two dimensions. In this work we employ the proposed characterization of interference to design an estimation algorithm which caters purely to a single dimension \emph{i.e.}, the statistical aspects of interference, and offers a more reliable metric than TID. 
\subsection{A Statistical Interference Estimation Approach}
 We propose a scheme predicated on the notion of \textit{statistical evenness} of channel allocation, which postulates that a proportionate distribution of the available channels among the radios will occasion an efficient CA. An even distribution of channels among radios in a WMN ensures fairness and boosts performance, as demonstrated in \cite{Manas2}. We name it the \textit{Channel Distribution Across Links} or \textit{CDAL} algorithm. The name is indicative of the underlying technique in which we determine the number of links operating on each channel taking the channel allocation to radios in the WMN as input.
 \begin{figure}[htb!]
                \centering 
                \includegraphics[width=7.5cm, height=1.7cm]{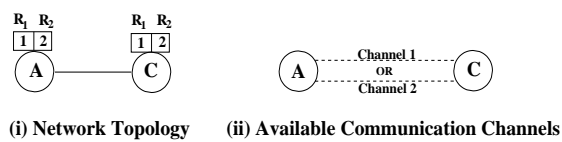}
                \caption{Link Selection For Transmission}
                \label{links}
        \end{figure}

A theoretical estimation approach is fundamentally static, and will fail to acknowledge the \textit{dynamic} or \textit{temporal} characteristics of interference. Further, determining the link selected for a radio transmission and identifying the channel assigned to the link are non-trivial problems, as link selection is a Media Access Control (MAC) mechanism. The standard approach of \textit{one flow transmission per radio} mandates that the radio which experiences the least interference or exhibits the highest \textit{signal to interference plus noise ratio} (SINR) should be used for the transmission. Thus, the link with the best network parameters is selected for transmission. But the \textit{quality} of a link vis-\`{a}-vis the interference degrading its efficiency, is a dynamic or temporal entity and can only be observed in real-time data transmissions. Also, the concept of \textit{parallel transmissions} is being leveraged in wireless networks by transmitting data simultaneously through 
multiple radios installed on a node \cite{Parallel}. These factors further complicate the theoretical determination of the link over which a transmission may occur in real-time.

To overcome these constraints, we adopt a \textit{probabilistic link selection} approach. In doing so, we account for the temporal characteristics up to some extent by introducing randomness in the link determination. Let us consider the trivial network topology presented in Figure~\ref{links}~(i), where both node $A$ and node $C$ are equipped with two identical radios. Let $Channel_1$ and $Channel_2$ be two orthogonal channels which are assigned to one radio each of both the nodes. If nodes $A$ and $C$ wish to communicate with each other at any moment, they can use either of the two non-conflicting links \emph{i.e.}, wireless $Link_1$ over $Channel_1$ or wireless $Link_2$ over $Channel_2$. It is difficult to ascertain the temporal selection made at the MAC layer. Thus, we take a probabilistic view  that if the described two-node network were to actively transmit data for an infinite period, $Link_1$ and $Link_2$ would be equally likely to be selected for transmission, as the probability of either link being 
chosen would converge to $1/2$. From a perspective of practical application, we invoke the \textit{central limit theorem} and assume that in case of availability of multiple links for transmission, each is equally likely to be chosen. Results will demonstrate that this innovative link selection approach leads to reliable estimates, because it facilitates the inclusion and accounting of the temporal characteristics of interference. 

\renewcommand{\algorithmicrequire}{\textbf{Input:}}
\renewcommand{\algorithmicensure}{\textbf{Output:}}
\begin{algorithm}[htb!] 
\caption{Channel Distribution Across Links}
\label{CDAL}
\begin{algorithmic}[1]
{\fontsize{9}{10}
\REQUIRE $G = (V,E)$, $R_i (i \in V)$, $CA = \{(R_i,C), i \in V\}$,\\ $CS =\{1, 2,...M\}$ \\
\textit{Notations} $:$ $G$ $\leftarrow$  WMN Graph, $R_i$ $\leftarrow$ Radio-Set,\\
		   $CS$ $\leftarrow$ Available Channel Set
\ENSURE CDAL$_{cost}$ \\
\line(1,0){236}
\STATE $LS \leftarrow FindLinkSet(G,CA)$ \COMMENT {$LS$ : Set of all wireless links present in $G$}
\STATE $CD \leftarrow ProbChannelSelect()$ \COMMENT {$CD$ : Set of link-count of each channel in $CS$}
\STATE $CDAL_{cost}\leftarrow StdDev(CD)$

\STATE \textit{Output CDAL$_{cost}$} \COMMENT{Estimation Metric of CDAL algorithm}
}
\end{algorithmic}
\end{algorithm}

\renewcommand{\algorithmicrequire}{\textbf{Input:}}
\renewcommand{\algorithmicensure}{\textbf{Output:}}
\begin{algorithm}[htb!] 
\caption{Function ProbChannelSelect()\\ (Probabilistic Selection Of Links)}
\label{probsel}
\begin{algorithmic}[1]
{\fontsize{9}{10}
\REQUIRE$G = (V,E)$, $CS =\{1, 2,...M\}$, $CD[M]$ \\
 $G$ : WMN Graph, $CS$ : Available Channel Set\\
 $CD$ : Channel Distribution Set of size $M$. 
\ENSURE Output $CD$ \emph{i.e.}, link count of each channel in $CS$ \\
\line(1,0){236}
\FOR {$i \in V$}
\STATE Determine $Ch_i$ and $Adj_i$ 
\COMMENT {$Ch_i$  : Set of channels allocated to the radios at node $i$ in $G$.  $Adj_i$ : Set of nodes adjacent to node $i$ in $G$}
\ENDFOR

\FOR {$i \in V$}
\FOR {$j \in Adj_i$}
\STATE $Get$ $ComCh_{ij}$ \COMMENT {Set of common channels assigned to radios of nodes $(i \ \& \ j)$}
\STATE let $p \leftarrow |ComCh_{ij}| $
\FOR {$k \in ComCh_{ij}$}
\STATE $CD[k-1] \leftarrow CD[k-1]+(k/p)$
\COMMENT {Increment the link count of channel $k$}
\ENDFOR
\ENDFOR
\ENDFOR
}
\end{algorithmic}
\end{algorithm}


We now the present the CDAL algorithm followed by a detailed description and a suitable example. The CDAL algorithm, illustrated in Algorithm~\ref{CDAL}, generates an interference estimate for CA schemes which we call the $CDAL_{cost}$. The first step in the process is to determine the set of all wireless links present in $G$, which is accomplished by the function $FindLinkSet$. The next step entails a probabilistic selection of links, followed by the task of ascertaining the distribution of channels across all links. These tasks are performed by the function $ProbChannelSelect$ which implements Algorithm~\ref{probsel}. For every channel, the function computes the \textit{link-count} \textit{i.e.}, the number of links that have been assigned that particular channel and inserts the link-count value into the set $CD$. Therefore the cardinality of set $CD$ is equal to the number of available channels \emph{i.e.}, the cardinality of set $CS$. A pair of nodes may have multiple channels at their disposal to 
communicate with each other. As per our assumption, each channel is equally likely to be selected and due to this probabilistic link selection approach, the link-count for a channel could be fractional.

Now, we devise a statistical mechanism to estimate the efficiency of a CA scheme. Translating the channel distribution into a statistical metric would require the processing of link-counts of channels on a purely quantitative basis. For example, in a 20 node WMN the ordered set $CD$ may have the link-count elements \{9,8,6\} for the three non-overlapping channels $C_1$, $C_2$ and $C_3$, respectively. We contend that a $CD$ of \{6,9,8\} or \{8,9,6\} should generate the same final CDAL$_{cost}$ as a $CD$ of \{9,8,6\}, because we are observing the channel-distribution with a purely statistical perspective. 

To engineer a quantitative statistical metric, we compute the \textit{standard deviation} of link-counts in $CD$ and consider it to be the CDAL$_{cost}$. There is a two-fold objective in employing the standard deviation of link-counts as the metric. First, it is a measure of the variation or dispersion of a set of data values from the mean. Since an equitable distribution of channels among radios is desirable, the closer CDAL$_{cost}$ is to 0, the more proportionate is the channel allocation. Second, as the size of WMN is scaled up in terms of nodes and the number of radios, TID computation becomes more complex and computationally intensive. In contrast, determining CDAL$_{cost}$ of a CA requires lesser computational overhead. Since the CDAL$_{cost}$ is a measure of dispersion from the ideal equitable distribution, lower is the magnitude of the CDAL$_{cost}$, better is the expected performance of CA when deployed in a WMN. 

We now illustrate the effectiveness of the CDAL estimation by applying it to two CA schemes, the \textit{Maximal Independent Set CA} (MIS) \cite{24Aizaz} and \textit{Radio Co-location Aware Optimized Independent Set CA} (OIS) \cite{Manas2}. OIS alleviates the radio co-location interference and incorporates \textit{statistical evenness} as a fundamental design objective. It is demonstrated in \cite{Manas2}, that OIS  outperforms MIS in terms of network performance metrics namely, throughput, packet loss ratio and mean delay. We compute the  CDAL$_{cost}$ for all the OISs and MISs implemented in grid WMNs of size $(N\times N)$, where $N\in \{5,\ldots,8\}$. The results are depicted in Table~\ref{MvsO}. It is evident that OIS CDAL$_{cost}$ is consistently lower than MIS CDAL$_{cost}$, which conforms to their relative performance in actual experiments. 

\begin{table} [h!]
\caption{OIS CDAL$_{cost}$ vs MIS CDAL$_{cost}$}
\tabcolsep=0.11cm
\begin{tabular}{|M{1.75cm}|M{1.75cm}|M{1.75cm}|M{2.5cm}|}
\hline 
 \multicolumn{1}{|c|}{\textbf{Grid}}&\multicolumn{1}{|c|}{\textbf{Num of}}&\multicolumn{2}{c|}{\textbf{CDAL$_{cost}$}}\\\cline{3-4}
     \multicolumn{1}{|c|}{\textbf{Size}}&\multicolumn{1}{|c|}{\textbf{Radios}}&\textbf{MIS}&\textbf{OIS}\\
\hline  
5$\times$5&50&4.48&2.86\\
\hline  
6$\times$6&72&6.86&6.33\\
\hline  
7$\times$7&98&8.87&5.88\\
\hline
8$\times$8&128&13.76&8.59\\
\hline  
\end{tabular} 
\label{MvsO}
\end{table}

\subsection{Time Complexity of CDAL Algorithm}

We consider an arbitrary MRMC WMN $G=(V,E)$ comprising of $n$ nodes, where each node is equipped with $m$ identical radios. The most computationally intensive step in the CDAL algorithm is to determine the $ComCh_{ij}$ for each pair of adjacent nodes $i\ \& \ j$, and then incrementing the link-count accordingly. This step has the worst case complexity of O($n\textsuperscript2 m\textsuperscript2$) which also makes it the upper bound for overall computational cost of the CDAL algorithm. In comparison, the complexity of determining the TID estimate is of the order O($n\textsuperscript2 m\textsuperscript3$). This is because the adjacency relationships have to be established at the radio-to-radio granularity, for all nodes in $G$. Next, conflict links are determined for each radio-to-radio link in the WMN, leading to an algorithmic complexity of O($n\textsuperscript2 m\textsuperscript3$). 
Thus, in comparison to the worst case complexity of TID estimate, CDAL algorithm fares better and this benefit will be more pronounced with the increase in the number of radios attached to a node.

\section{Simulations, Results and Analysis}
We assess the efficiency and performance of CDAL$_{cost}$ in comparison to the TID estimates through a meticulous procedure elucidated here. \textit{(a)} Choose the WMN topology and a comprehensive data traffic scenario. \textit{(b)} Implement a heterogeneous mix of CA schemes. \textit{(c)} Run extensive simulations to obtain aggregate performance metrics for the CAs. \textit{(d)} Subject the CAs to the two interference estimation approaches \emph{viz.}, TID and CDAL$_{cost}$. \textit{(e)} Consider the sequence of CAs with respect to observed performance metrics as reference, and determine the \textit{error in sequence} in each of the estimation approaches.

\subsection{Simulation Parameters}
Simulations are performed in ns-3 \cite{NS-3} to record the performance of CAs in a simulated $5\times5$ grid WMN comprising of $25$ nodes. The simulation parameters are presented in Table~\ref{sim}. A 10 MB file is transmitted from the source to the destination in every multi-hop TCP and UDP flow. TCP and UDP transport layer protocols are implemented in \mbox{ns-3} through the inbuilt applications, \textit{BulkSendApplication} and \textit{UdpClientServer}, respectively. Through TCP simulations we determine the \textit{aggregate network throughput} while UDP simulations offer us the \textit{packet loss ratio}, which we henceforth denote as Throughput and PLR, respectively, for an easy discourse. 
\begin{table} [h!]
\caption{ns-3 Simulation Parameters}
   \center 
\begin{tabular}{|p{5cm}|p{3cm}|}
\hline
\bfseries
 Parameter&\bfseries Value \\ [0.2ex]
 \hline
\hline
No. of Radios/Node&2   \\
\hline
Range Of Radios&250 mts   \\
\hline
Available Orthogonal Channels&3 in 2.4 GHz  \\
\hline
Maximum 802.11g PHY Datarate &54 Mbps  \\
\hline
Maximum Segment Size (TCP)&1 KB   \\
\hline
Packet Size (UDP)&1KB\\
\hline
MAC Fragmentation Threshold&2200 Bytes  \\
\hline
RTS/CTS &Enabled  \\
\hline
Routing Protocol &OLSR    \\
\hline
\end{tabular}
\label{sim}
\end{table}     

\subsection{Traffic Characteristics and Test Scenarios}
We conceive a comprehensive set of \textit{data traffic characteristics} which is crucial to highlight the performance bottlenecks created by the endemic interference. We establish \mbox{4-Hop-Flows} between the first and the last node of each row and column, and \mbox{8-Hop-Flows} between the diagonal nodes of the grid. Various combinations of these multi-hop flows are formulated to engineer test-scenarios for the grid WMN. For each test-case two set of experiments are carried out \emph{viz.}, one set employs only TCP flows and the other comprises of only UDP flows. Four high traffic test-scenarios consisting of the following number of concurrent multi-hop flows are simulated in the $25$ node grid WMN :\\ 
(i) $5$ \quad \quad  (ii) $8$ \quad \quad (iii) $10$  \quad  \quad (iv) $12$. \\
These test-scenarios of increasing levels of interference adequately capture the interference characteristics of a WMN, and are thus ideal to demonstrate the overall performance of a CA implemented in a WMN. 
 
\subsection{Selection of CA Schemes} 
We implement a diverse set of 5 CA schemes, which range from the low performance \textit{centralized static CA} scheme (CEN) \cite{23Cheng} to the high performance \textit{grid specific CA} scheme (GSCA). GSCA is designed for maximal performance in the current simulation set-up by ensuring a minimum TID of the channel allocation through a rudimentary brute-force approach. Other CA schemes that we implement are the \textit{breadth first traversal} approach (BFS) \cite{22Ramachandran}, a static \textit{maximum clique} based algorithm (CLQ) \cite{17Xutao} and a \textit{maximum independent set} based scheme (MIS) \cite{24Aizaz}. 

We employ two multi-radio multi-channel conflict graph models (MMCGs) \emph{viz.}, the conventional MMCG (C-MMCG) and the enhanced MMCG (E-MMCG) \cite{Manas} to implement each of the above CA schemes, except GSCA. C-MMCG represents link conflicts in a traditional fashion and does not take into consideration the impact of radio co-location interference (RCI) in a wireless network. E-MMCG is an improved version of its conventional counterpart as it does a comprehensive accounting of all RCI interference scenarios in its link conflict representation of the WMN. CAs that are generated with E-MMCG as the underlying conflict graph model demonstrate more effective interference mitigation which results in an enhanced network performance \cite{Manas}. Thus, we have a total of 9 CA schemes and for ease of reference, we will denote a C-MMCG based CA as $CA_C$ and its corresponding E-MMCG version as $CA_E$. GSCA is denoted simply as $GSCA$.

\subsection{Results and Analysis}
\begin{figure}[htb!]
                \centering
                \includegraphics[width=8.5cm, height=6cm]{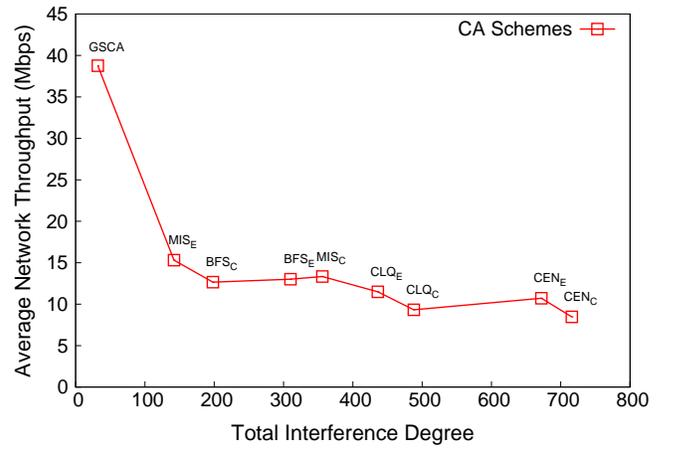}
                \caption{Observed Correlation of Avg Throughput with TID} 
                \label{TIDX1}
        \end{figure}

\begin{figure}[htb!]
	    \centering
	    \includegraphics[width=8.5cm, height=6cm]{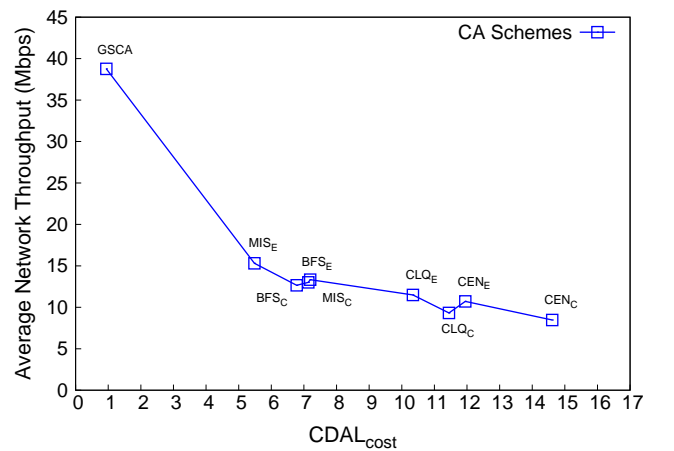}
	    \caption{Observed Correlation of Avg Throughput with CDAL$_{cost}$} 
	    \label{TIDX2}
    \end{figure}

\begin{figure}[htb!]
                \centering
                \includegraphics[width=8.5cm, height=6cm]{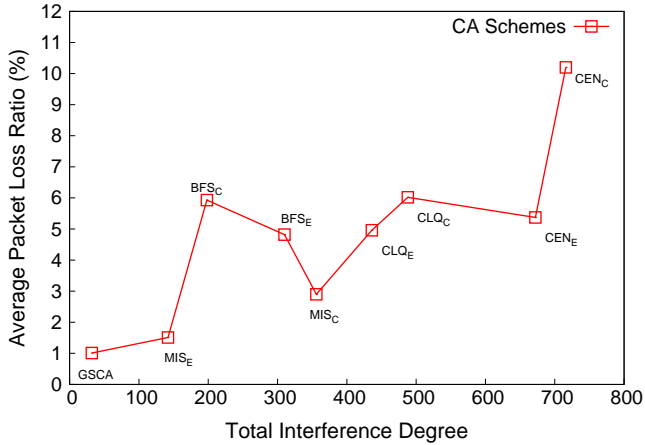}
                \caption{Observed Correlation of Avg PLR with TID} 
                \label{TIDX3}
        \end{figure}

\begin{figure}[htb!]
	    \centering
	    \includegraphics[width=8.5cm, height=6cm]{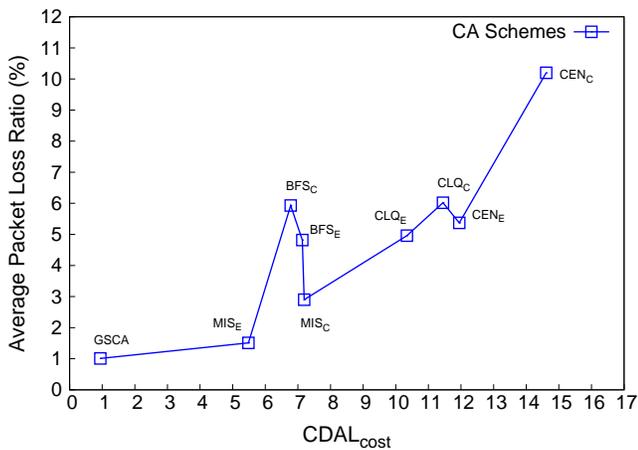}
	    \caption{Observed Correlation of Avg PLR with CDAL$_{cost}$} 
	    \label{TIDX4}
    \end{figure}

An exhaustive set of simulations were run and the values of performance metrics \emph{viz.}, Throughput and PLR were recorded. For each CA, we compute the mean of the recorded values of all test-scenarios to generate the average performance metrics for the CA, denoted by \textit{Avg Throughput} and \textit{Avg PLR}. For each CA, the TID estimate and CDAL$_{cost}$ are computed and plotted against the observed performance metrics. The results are presented in Figures~\ref{TIDX1}, \ref{TIDX2}, \ref{TIDX3} and \ref{TIDX4}.
We process the results by first ordering the CAs in a \textit{sequence} of increasing magnitude of each of the recorded performance metrics. Thereafter, we orient the CAs in increasing order of expected performance, as predicted by the two theoretical interference estimation metrics. Now that we have the CA sequences based on both, actual performance data and theoretical interference estimates, we compare the actual and theoretical CA sequences to determine the \textit{degree of confidence} (DoC) for the two CA performance prediction approaches. DoC reflects the efficacy of the interference estimation metric in predicting with high confidence, how a particular CA scheme will perform when implemented in a given WMN.

To determine the DoC, we first ascertain the \textit{error in sequence} (EIS) for each estimation metric. In an ordered sequence of $n$ CAs, a total of  $\textsuperscript n C_2$ comparisons can be made between individual CAs with respect to the magnitude of the metric in context. Considering the sequence of CAs determined by experimental metric values as reference, we ascertain how many \textit{comparisons} are in error in the sequences based on theoretical metrics. An erroneous comparison signifies that the performance relationship predicted by the estimation metric is contrary to the observed real-time performance. The sum of all erroneous comparisons in the CA sequence of an estimation metric is its EIS. Further, we compute the DoC of the estimation metric through the relation, \mbox{$DoC = (1-(EIS/\textsuperscript n C_2))\times100$}, where $n$ is the number of CAs in the sequence.

From the illustrated results, the CA sequence in terms of increasing Avg Throughput can be determined as : \textit{(CEN$_C$ $<$ CLQ$_C$ $<$ CEN$_E$ $<$ CLQ$_E$ $<$ BFS$_C$ $<$ BFS$_E$ $<$ MIS$_C$ $<$ MIS$_E$ $<$ GSCA)}, CEN$_C$ being the least efficient and GSCA being the best in the CA sample set. The CA sequences in terms of TID and CDAL$_{cost}$ can also be determined in similar fashion. For both these metrics, a high magnitude of the estimate implies high interference in the WMN \emph{i.e.}, ($CA \ Performance \propto 1/Estimate \ Value)$. Thus, we order the CAs in the decreasing order of estimate values. CA sequence for TID is : \textit{(BFS$_E$ $<$ CLQ$_C$ $<$ MIS$_E$ $<$ BFS$_C$ $<$ CEN$_E$ $<$ CEN$_C$ $<$ CLQ$_E$ $<$ MIS$_C$ $<$ GSCA)}. The CDAL$_{cost}$ CA sequence is :  \textit{(CEN$_C$ $<$ CEN$_E$ $<$ CLQ$_C$ $<$ CLQ$_E$ $<$ MIS$_C$ $<$ BFS$_E$ $<$ BFS$_C$ $<$ MIS$_E$ $<$ GSCA)}. Coming to CA comparisons, CDAL$_{cost}$ sequence causes only 4 upsets while TID registers as many as 15 false 
comparisons. Thus, in terms of Avg Throughput the EIS for CDAL$_{cost}$ and TID is 4 and 15, respectively. The corresponding DoC values which represent the measure of reliability of a prediction estimate, for a total of 36 ($\textsuperscript 9 C_2$) CA performance comparisons, are 88.89\% and 58.33\% for CDAL$_{cost}$ and TID, respectively. 

    
%
Performance evaluation of the theoretical interference estimation approaches, based on the observed results of the two metrics is presented in Table~\ref{DOC}. CDAL$_{cost}$ consistently outperforms TID estimate in terms of reliability of CA performance prediction and its accuracy levels stay above 80\%. Another positive feature of CDAL$_{cost}$ predication is that the best performing CA \emph{i.e}, GSCA and the least efficient CA \emph{i.e.}, CEN$_C$, are rightly predicted to be the best and the worst CA, respectively. TID estimates predict GSCA to be the best, which is substantiated by experimental results but wrongly predict BFS$_E$ to be the worst CA for the given grid WMN, although it outperforms the worst performing CA CEN$_C$ by over $56\%$ in terms of Avg Throughput. A qualitative assessment of Figures~\ref{TIDX1}, \ref{TIDX2}, \ref{TIDX3} and \ref{TIDX4}, will reveal that CDAL$_{cost}$ estimates succeed in differentiating between CAs that perform well (\emph{e.g.}, GSCA), CAs that exhibit an average 
performance (\emph{e.g.}, BFS$_E$) and CAs that are not suitable for the given WMN (\emph{e.g.}, CEN$_C$). In sharp contrast, TID estimates fail to effect this accurate distinction between CAs.
\begin{table} [h!]
\caption{Performance Evaluation Of Estimation Metrics}
\tabcolsep=0.10cm
\begin{tabular}{|M{2.5cm}|M{1.2cm}|M{1.4cm}|M{1.2cm}|M{1.4cm}|}
\hline 
\multicolumn{1}{|c|}{\textbf{Performance}}&\multicolumn{2}{|c|}{\textbf{TID}}&\multicolumn{2}{|c|}{\textbf{CDAL$_{cost}$}}\\\cline{2-5}
\multicolumn{1}{|c|}{\textbf{Metric}}&\multicolumn{1}{|c|}{\textbf{EIS}}&\multicolumn{1}{|c|}{\textbf{DoC} (\%)}&\multicolumn{1}{|c|}{\textbf{EIS}}&\multicolumn{1}{|c|}{\textbf{DoC (\%)}}\\
\hline  
Avg Throughput&15&58.33&4&88.89\\
\hline 
Avg PLR&12&66.67&7&80.55\\
\hline 
\end{tabular} 
\label{DOC}
\end{table}

\section{Conclusions}
The motive of our work was to engineer an interference estimation algorithm which assures greater adherence to actual results and offers a reliable metric to assist in the task of CA selection for a given WMN. The results have demonstrated that CDAL$_{cost}$ has met these objectives and is decidedly a better metric than TID at a lesser computational overhead. The efficacy of CDAL$_{cost}$ has also strengthened the proposed qualitative characterization of interference endemic in WMNs as its algorithmic design is motivated by this characterization.
\section{Future Work}
We intend to devise a metric that also takes into account the \textit{spatial} characteristics of interference in addition to the statistical aspects. We also plan to perform a quantitative analysis of these estimation metrics.

\bibliography{ref}

\begin{thebibliography}{10}
\providecommand{\url}[1]{#1}
\csname url@samestyle\endcsname
\providecommand{\newblock}{\relax}
\providecommand{\bibinfo}[2]{#2}
\providecommand{\BIBentrySTDinterwordspacing}{\spaceskip=0pt\relax}
\providecommand{\BIBentryALTinterwordstretchfactor}{4}
\providecommand{\BIBentryALTinterwordspacing}{\spaceskip=\fontdimen2\font plus
\BIBentryALTinterwordstretchfactor\fontdimen3\font minus
  \fontdimen4\font\relax}
\providecommand{\BIBforeignlanguage}[2]{{%
\expandafter\ifx\csname l@#1\endcsname\relax
\typeout{** WARNING: IEEEtran.bst: No hyphenation pattern has been}%
\typeout{** loaded for the language `#1'. Using the pattern for}%
\typeout{** the default language instead.}%
\else
\language=\csname l@#1\endcsname
\fi
#2}}
\providecommand{\BIBdecl}{\relax}
\BIBdecl

\bibitem{NPcomplete}
O.~D. Incel, A.~Ghosh, B.~Krishnamachari, and K.~K. Chintalapudi,
  ``Multi-channel scheduling for fast convergecast in wireless sensor
  networks,'' \emph{Department of Computer Science, University of Twente, Tech.
  Rep}, 2008.

\bibitem{Ding}
Y.~Ding and L.~Xiao, ``Channel allocation in multi-channel wireless mesh
  networks,'' \emph{Computer Communications}, vol.~34, no.~7, pp. 803--815,
  2011.

\bibitem{TID1}
S.~Hoteit, S.~Secci, R.~Langar, and G.~Pujolle, ``A nucleolus-based approach
  for resource allocation in ofdma wireless mesh networks,'' \emph{Mobile
  Computing, IEEE Transactions on}, vol.~12, no.~11, pp. 2145--2154, 2013.

\bibitem{Arunabha}
A.~Sen, S.~Murthy, S.~Ganguly, and S.~Bhatnagar, ``An interference-aware
  channel assignment scheme for wireless mesh networks,'' in
  \emph{Communications, 2007. ICC'07. IEEE International Conference on}.\hskip
  1em plus 0.5em minus 0.4em\relax IEEE, 2007, pp. 3471--3476.

\bibitem{Manas}
\BIBentryALTinterwordspacing
S.~M. Kala, M.~Reddy, R.~Musham, and B.~R. Tamma, ``Interference mitigation in
  wireless mesh networks through radio co-location aware conflict graphs,''
  2015. [Online]. Available: \url{http://dx.doi.org/10.1007/s11276-015-1002-4}
\BIBentrySTDinterwordspacing

\bibitem{26Koutsonikolas}
S.~M. Das, D.~Koutsonikolas, Y.~C. Hu, and D.~Peroulis, ``Characterizing
  multi-way interference in wireless mesh networks,'' in \emph{Proceedings of
  the 1st international workshop on Wireless network testbeds, experimental
  evaluation \& characterization}.\hskip 1em plus 0.5em minus 0.4em\relax ACM,
  2006, pp. 57--64.

\bibitem{Alignment}
L.~E. Li, R.~Alimi, D.~Shen, H.~Viswanathan, and Y.~R. Yang, ``A general
  algorithm for interference alignment and cancellation in wireless networks,''
  in \emph{INFOCOM, 2010 Proceedings IEEE}.\hskip 1em plus 0.5em minus
  0.4em\relax IEEE, 2010, pp. 1--9.

\bibitem{Manas2}
S.~M. Kala, M.~Reddy, R.~Musham, and B.~R. Tamma, ``Radio co-location aware
  channel assignment for interference mitigation in wireless mesh networks,''
  \emph{arXiv preprint arXiv:1503.04533}, 2015.

\bibitem{Parallel}
H.~Yun, Y.~Shoubao, Z.~Qi, and Z.~Peng, ``Parallel-transmission: a new usage of
  multi-radio diversity in wireless mesh network,'' \emph{Int'l J. of
  Communications, Network and System Sciences}, vol. 2009, 2009.

\bibitem{24Aizaz}
A.~U. Chaudhry, J.~W. Chinneck, and R.~H. Hafez, ``Channel requirements for
  interference-free wireless mesh networks to achieve maximum throughput,'' in
  \emph{Computer Communications and Networks (ICCCN), 2013 22nd International
  Conference on}.\hskip 1em plus 0.5em minus 0.4em\relax IEEE, 2013, pp. 1--7.

\bibitem{NS-3}
T.~R. Henderson, M.~Lacage, G.~F. Riley, C.~Dowell, and J.~Kopena, ``Network
  simulations with the ns-3 simulator,'' \emph{SIGCOMM demonstration}, 2008.

\bibitem{23Cheng}
H.~Cheng, G.~Chen, N.~Xiong, and X.~Zhuang, ``Static channel assignment
  algorithm in multi-channel wireless mesh networks,'' in \emph{International
  Conference on Cyber-Enabled Distributed Computing and Knowledge Discovery,
  2009. CyberC'09.}\hskip 1em plus 0.5em minus 0.4em\relax IEEE, 2009, pp.
  49--55.

\bibitem{22Ramachandran}
K.~N. Ramachandran, E.~M. Belding-Royer, K.~C. Almeroth, and M.~M. Buddhikot,
  ``Interference-aware channel assignment in multi-radio wireless mesh
  networks.'' in \emph{INFOCOM}, vol.~6, 2006, pp. 1--12.

\bibitem{17Xutao}
Y.~Xutao and X.~Jin, ``A channel assignment method for multi-channel static
  wireless networks,'' in \emph{2011 Global Mobile Congress}, 2011, pp. 1--4.

\end{thebibliography}

\end{document}